# Comment on: «Describing temperature-dependent self-diffusion coefficients and fluidity in 1- and 3-alcohols using the compensated Arrhenius formalism» [J. Phys. Chem. B 2016, 120, 37, 9959–68]

Alexander Kholmanskiy

Science Center «Bemcom», Moscow, Russia,

allexhol@ya.ru, http://orcid.org/0000-0001-8738-0189

In Supplemental Information (SI) to the article [A. M. Fleshman, G. E. Forsythe, M. Petrowsky, R. Frech, J. Phys. Chem. B 2016, 120, 37, 9959-68] the temperature dependences (TDs) of the dielectric and dynamic properties of polyhydric alcohols, as well as the activation energy ($E_a$) obtained using the compensated Arrhenius formalism (CAF) are given. The authors present CAF as an approach provides physical significance into the molecular level nature of the transport process. However, the CAF logic is physically inadequate and analysis of TDS polyhydric alcohols using CAF cannot provide physically significant correlations useful for understanding the molecular mechanisms of fluid dynamics. However, such correlations from SI can be obtained by applying physically adequate Arrhenius approximations of TDs properties of liquids [A. Kholmanskiy, N. Zaytseva, J. Mol. Liq. 275 (2019) 741-8].

*The CAF separates a given transport property's TD into two contributing factors: the exponential prefactor $D_0[\varepsilon_s(T)]$ and the Boltzmann factor $(\exp[-E_a/RT])$* [1]. However, the stationary dielectric constant ($\varepsilon_s$) is itself a derivative of the process of rotational-translational diffusion of the dipole moments of molecules ($\mu$). Therefore, TD $\varepsilon_s$ is determined by the TDs of the self-diffusion coefficient (D) and dynamic viscosity ($\eta$), and not vice versa. Since the $\mu$ values in both series of the studied alcohols are almost identical, the changes in the activation energies $E_a^\varepsilon$, $E_a^\eta$ and $E_a^D$ in the series of alcohol homologues should correlate in a certain way. The value of $E_a^\varepsilon$ in [1] was not determined, but taking into account the additional TD due to $\varepsilon_s$ contained in the exponential prefactor led to an increase in $E_a^\eta$ and $E_a^D$ by 20-40% compared to $E_a$ determined using a simple Arrhenius approximation (CAE) [1]. The difference between CAF $E_a^\eta$ and CAF $E_a^D$ for 1-alcohols in the article and Supplemental Information (SI) is also ~10-15% (see Table 1).

**Таблица 1**

Average energies of activation ($E_a$) for 1-alcohols and 3-alcohols calculated from self-diffusion coefficient (D) and fluidity (1/$\eta$) data in article [1] and Supplemental Information (SI)

| Average $E_a$ (kJ/mol) | 1-alcohol | | 3-alcohol | |
|---|---|---|---|---|
| | article | SI | article | SI |
| CAF $E_a^D$ | 42.3±1% | 35.9±1.4% | 43.4±2% | 43.5±0.9% |
| CAF $E_a^\eta$ | 36.5±1.4% | 32.9±0.6% | 34±2.9% | 35±2.8% |

It is clear that such a technology for calculating $E_a$ and the absence of $E_a^\varepsilon$ exclude the possibility of obtaining physically significant and reliable information on the dynamics of alcohols at the molecular level. However, such correlations $E_a^\varepsilon$, $E_a^\eta$ and $E_a^D$ between themselves (see Table 2) and the structural parameters of molecules can be established by analyzing TDs D, η, $\varepsilon_s$ and alcohol densities (ρ), given in SI using CAE:

$$F_A = \exp(\pm E_a/RT)$$

**Таблица 2**

Energies of activation ($E_a$) and correlation trends for 1-alcohols and 3-alcohols calculated from Arrhenius approximations TDs of self-diffusion coefficient (D), fluidity (1/η), dielectric constant ($\varepsilon_s$), and density (ρ). Initial data from Supplemental Information

| Fluid | | CAE $E_a$ (kJ/mol) | | | |
|---|---|---|---|---|---|
| | | $-E_a^D$ | $E_a^\eta$ | $E_a^\varepsilon$ | $E_a^\rho$ |
| 1- | hexanol | 27.5 | 22.2 | 7.3 | 0.77 |
| | heptanol | 27.9 | 23.6 | 7.5 | 0.74 |
| | octanol | 29.5 | 25.1 | 7.4 | 0.71 |
| | nonanol | 30 | 26 | 7.6 | 0.73 |
| | decanol | 30.5 | 27.5 | 7.7 | 0.74 |
| trend or mean | | $0.037m$ | $0.0079J$ | $0.054E_a^\eta$; $0.069E_a^D$ | $0.74\pm2\%$ |
| 3- | hexanol | 35.1 | 29.7 | 8.1 | 0.86 |
| | heptanol | 35.9 | 29.5 | 7.2 | 0.83 |
| | octanol | 36.2 | 29.2 | 5 | 0.83 |
| | nonanol | 36.7 | 28.6 | 3.6 | 0.85 |
| | decanol | 37.3 | 30.1 | 2.4 | 0.81 |
| trend or mean | | $0.06m$ | $29.4\pm1\%$ | $4.1E_a^\eta$; $-2.8E_a^D$ | $0.84\pm2\%$ |

The dependence of $E_a^\varepsilon$ on $E_a^D$ and $E_a^\eta$ is evidenced by the qualitative correlations in Figure 1 and Figure 2. The $E_a^D$ and $E_a^\eta$ values in the 1- and 3-alcohols series correlate with the molecular weight ($M_r$) and the moment of inertia about the axis passing through the middle of the molecule. The molecules were compared with a uniform rod of length $L$ and, neglecting the contribution of oxygen, the value of $M_r$ was taken to be proportional to the number of carbon atoms (k), and $L$ to the number of C-C (k-1) bonds, then J ~ k (k-1)$^2$. The strange inversion of the correlation $E_a^\varepsilon \propto E_a^D$ and the loss of the 3-decanol point from the correlation $E_a^\varepsilon \propto E_a^\eta$ are apparently caused by an experimental error or inadequacy of the technology for calculating $\varepsilon_s$ for homologues of 3-alcohols. The key parameter in these calculations is the T-dependent «dipole density» (N(T)):

«*The relationship between N(T) and $\varepsilon_s$ is an important aspect of the CAF. The N(T) is calculated by dividing the solution density by the molecular weight*» [1]: $N(T)=\rho/M_r$.

As follows from Table 2, the value of $E_a^\rho$ in both series of alcohol homologues remains constant within the measurement error. Hence it follows that the value $E_a^N$ should not change in the series of homologues, which confirms the independence of the values CAF $E_a^D$ and CAF $E_a^\eta$ on the structure of homologues. The physical inadequacy of this result is characteristic of CAF.

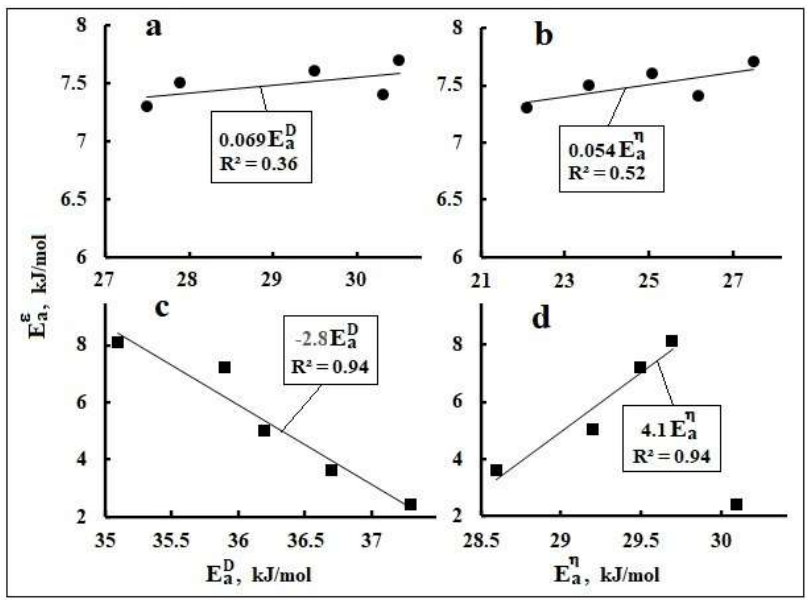

Figure 1. Dependences of the activation energy of the dielectric constant ($E_a^\varepsilon$) on the activation energies of the self-diffusion coefficient ($E_a^D$) and viscosity ($E_a^\eta$) for homologues of 1-alcohols (**a, b**) and 3-alcohols (**c, d**) ). Activation energies are calculated using the CAE method. Initial data from Supplemental Information [1].

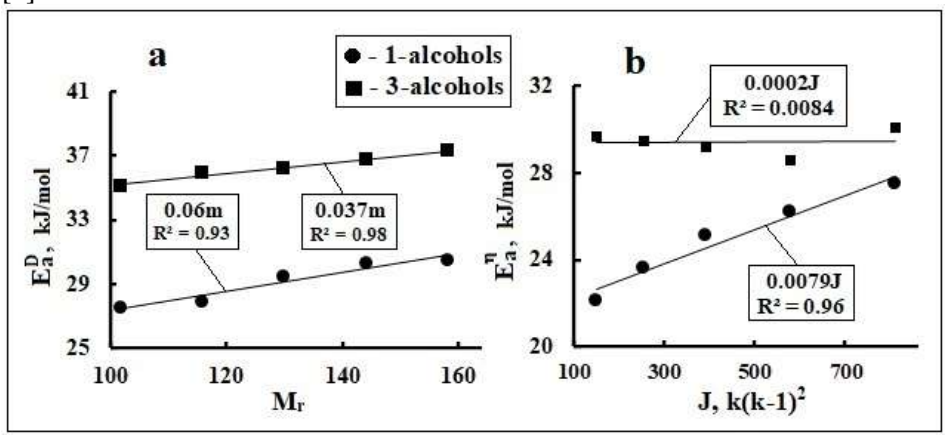

Figure 1. Dependences of the activation energies of the self-diffusion coefficient ($E_a^D$) on the molecular weight of homologues of 1- and 3-alcohols (**a**) and viscosity ($E_a^\eta$) on the moment of inertia of homologues of 1- and 3-alcohols containing k carbon atoms (**b**) Activation energies are calculated using the CAE method. Initial data from Supplemental Information [1].

A physically adequate modification of CAE for the dielectric and dynamic properties of liquids is the representation of $F_A$ by a bimodal function [2]:

$$F_A = \exp(\pm E_a/RT) = T^{\pm\beta} \exp(\pm E_R/RT), \qquad (1)$$

where $T^{\pm\beta}$ c $\beta = 0, \pm\frac{1}{2}, \pm 1$ corresponds to the exponential prefactor and $E_R$ represents the activation energy of the reaction, which is mainly responsible for the TD characteristics of the liquid. The $\beta$ values are selected taking into account the equations known in molecular physics (Stokes-Einstein-Debye, Kirkwood, Clapeyron). For D, $\eta$, $\rho$ and $\varepsilon_s$, the $\beta$ values are: 1, 0, -1 and -1, respectively [2]. To separate $E_R$ from $E_a$, the function $T^{\pm\beta}$ in [2] was presented as $\exp(\pm E_T/RT)$ and obtained:

$$\pm E_a = \pm E_R \pm E_T. \qquad (2)$$

The $E_T$ value for the T series in [1] is 2.7 kJ/mol for D, -2.7 kJ/mol for $\rho$ and $\varepsilon_s$, and 0 for $\eta$. Taking into account these values from the formula (2) and the data of Table 2, for example, for 1-decanol, such $E_R$ values for D, $\eta$, $\rho$ and $\varepsilon_s$ follow (in kJ/mol): -27.8; 27.5; -1.91 and 5.0, respectively. The minus signs of $E_R$ correspond to endothermic reactions, and plus – exothermic, in which energy is released when the hydrogen bonds are rearranged during the rotation of the molecule. Due to this energy, a water molecule, for example, can free itself from hydrogen bonds and make a translational jump [3, 4].

The key question of molecular dynamics of alcohols is: what are the structure and dynamics of the hydrogen bonding network? In [1], only one phrase applies to this problem: *«It is known that 3-alcohols exhibit less extensive hydrogen bonding networks than 1-alcohols»*, which is given without reference to the source. Meanwhile, the application of a physically adequate approximation (1) to TDs of the characteristics of alcohols given in SI showed that for them, similar to TDs in the properties of water [2], a kink is found in the vicinity of 298 K. At T <298 K, $E_a$ significantly increases, since the role of supramolecular structures made of hydrogen bonds increases. For example, for 3-decanol $E_a^D$, $E_a^\eta$ and $E_a^\varepsilon$ in T intervals of 5-25 °C and 35-85 °C were in kJ/mol: -43.1 and -34.5; 38.4 and 27.1; 4.1 and 1.8, respectively.

In conclusion, it should be said that the physical inadequacy of the CAF technology can also affect the level of reliability of the results of studies of the dynamics of polar liquids and electrolytes, published in J. Phys. Chem. B 2009, 113, 5996-6000; 2010, 114, 8600-8605; 2012, 116, 10098-10105.